\documentclass[twocolumn,prb,floatfix,amsmath,amssymb,showpacs]{revtex4-1}
\setcitestyle{numbers,square}
\usepackage{multirow}
\usepackage{amsfonts}%
\usepackage{mathrsfs,bbm}%
\usepackage{xcolor}
\usepackage{graphicx}%
  \DeclareGraphicsExtensions{.pdf,.eps}
\usepackage{prettyref}%
	\newcommand{\pref}[1]{\prettyref{#1}}%
	\newrefformat{fig}{Fig.~\ref{#1}}%
	\newrefformat{tab}{Table~\ref{#1}}%
	\newrefformat{sec}{Sec.~\ref{#1}}%
	\newrefformat{eq}{Eq.~(\ref{#1})}%

\newcommand{\qe}{\textsc{Quantum~ESPRESSO}}
\newcommand{\adeg}[1]{\ensuremath{#1^{\circ}}}
\newcommand{\bfk}{\ensuremath{\textbf{k}}}%
\newcommand{\bfa}{\ensuremath{\textbf{a}}}%
\newcommand{\bfb}{\ensuremath{\textbf{b}}}%
\newcommand{\bfc}{\ensuremath{\textbf{c}}}%
\newcommand{\ang}{\ensuremath{\textnormal{ \AA}}}
\newcommand{\ev}{\ensuremath{{\textrm{ eV}}}}%
\newcommand{\mev}{\ensuremath{{\textrm{ meV}}}}%
\newcommand{\ef}{\ensuremath{E_{\mathrm{F}}}}%
\newcommand{\etot}{\ensuremath{E_{\mathrm{tot}}}}%
\newcommand{\dxz}{\ensuremath{d_{xz}}}%
\newcommand{\dyz}{\ensuremath{d_{yz}}}%
\newcommand{\dxy}{\ensuremath{d_{xy}}}%
\newcommand{\pbnm}{\ensuremath{Pbnm}}%
\newcommand{\ptobm}{\ensuremath{P2_1/m}}%
\newcommand{\ptobb}{\ensuremath{P2_1/b}}%
\newcommand{\lto}{LaTiO$_3$}%
\newcommand{\lvo}{LaVO$_3$}%
\newcommand{\sto}{SrTiO$_3$}%
\newcommand{\ttg}{\ensuremath{t_{2g}}}
\newcommand{\asub}{\ensuremath{a_{\mathrm{sub}}}}
\newcommand{\acub}{\ensuremath{a_{\mathrm{cub}}}}
\newcommand{\atet}{\ensuremath{a_{\mathrm{tetr}}}}
\newcommand{\coop}{\ensuremath{c_{\mathrm{oop}}}}
\newcommand{\dvo}{\ensuremath{d_{\rm V-O}}}
\newcommand{\trot}[1]{\ensuremath{\phi^{#1}_{\rm V\hat{O}V}}}
\newcommand{\umit}{\ensuremath{U_{\textrm{\sc mit}}}}

\begin{document}

\title{Structural and electronic properties of epitaxially-strained LaVO$_3$ from density functional theory and dynamical mean-field theory}
\date{\today}
\author{Gabriele Sclauzero}
\email{gabriele.sclauzero@mat.ethz.ch}
\author{Claude Ederer}
\affiliation{%
Materials Theory, ETH Zurich, Wolfgang-Pauli-Strasse 27, CH-8093 Z\"u{}rich, Switzerland}
\pacs{71.30.+h,71.15.−m,73.22.−f}
\keywords{perovskites, thin films, epitaxial strain, metal-insulator transition}
\begin{abstract}
The effect of epitaxial strain on the structural and electronic properties of \lvo\ is investigated through density functional theory (DFT) and dynamical mean field theory (DMFT).
Two different growth orientations of the crystal are considered, one preserving the bulk \pbnm\ space-group symmetry and another giving rise to a symmetry lowering to \ptobm.
In the nonmagnetic DFT structures, the two growth orientations are equally favored for all tensile strains considered here, as well as for compressive strains weaker than $-3\%$.
For stronger compressive strains, the \ptobm\ orientation is favored and shows a complete suppression of octahedral tilts along the out-of-plane direction.
Magnetically-ordered structures do not show a complete tilt suppression, but the trend points to a similar reduction of the out-of-plane V--O--V bond angles under compressive strain.
Our DMFT calculations show that, in accord with room-temperature experiments, the bulk paramagnetic Mott-insulating state of \lvo\ is robust against epitaxial strains attainable in thin films, since the suppression of orbital fluctuations counteracts the effect of bandwidth increase with compressive strain.
Under stronger compressive strains, the straightening of the V--O--V bonds in the \ptobm\ geometry interferes with the suppression of orbital fluctuations and hence perturbs the Mott phase more strongly, albeit not enough to achieve a metallic phase.
\end{abstract}

\maketitle

\section{Introduction}
\label{sec:intro}

Transition metal oxides are very interesting because they host complex phenomena, such as one or more ferroic orderings, often coupled with each other, Jahn-Teller distortions, magnetic phase transitions, superconductivity, and more \cite{Dagotto2005Science}.
The electronic structure of these materials is often characterized by strong electron-electron correlations, mainly because of the localized nature of the $d$ electrons of the transition metal cations \cite{Imada1998RMP}.
An important class of complex oxides are the early-3$d$ transition-metal perovskites, $AB$O$_3$, where $B$ is a transition metal cation with $d^1$ or $d^2$ electron configuration, such as Ti$^{3+}$, V$^{4+}$, V$^{3+}$, or Cr$^{4+}$. 
Here, the correlated electron manifold results from the three nearly-degenerate \ttg\ states, which are split off from the two energetically higher $e_g$ orbitals by the cubic component of the crystal field.
Hybridization of the \ttg\ orbitals with the $p$ states of the oxygen ligands then leads to the formation of partially-filled antibonding bands with predominant transition metal \ttg\ orbital character.

Progress in growth techniques has opened a new avenue for tailoring complex oxide heterostructures with desired properties \cite{Mannhart2010Science,Hwang2012NMat}.
Possible engineered heterostructures include thin films as well as superlattices, where layers of different oxides are periodically stacked.
In both cases, a two-dimensional stress field, due to the lattice mismatch between the different constituents, as well as one or more interfaces arise within these structures.
The resulting epitaxial strain, as well as the interface effects, can potentially alter the electronic properties of such heterostructures.
For instance, the polar discontinuity at the LaAlO$_3$/\sto\ interface induces a high-mobility two-dimensional electron gas at the interface of these otherwise insulating materials \cite{Ohtomo2004Nat}. 
Also the \lvo/\sto\ Mott-insulator/band-insulator heterointerface shows an insulator-to-metal transition, attributed to the $n$-type VO$_2$/LaO/TiO$_2$ polar discontinuity \cite{Hotta2007PRL}, indicating interface conduction arising from electronic reconstructions.

\begin{figure}[b]
  \includegraphics[width=8.6cm]{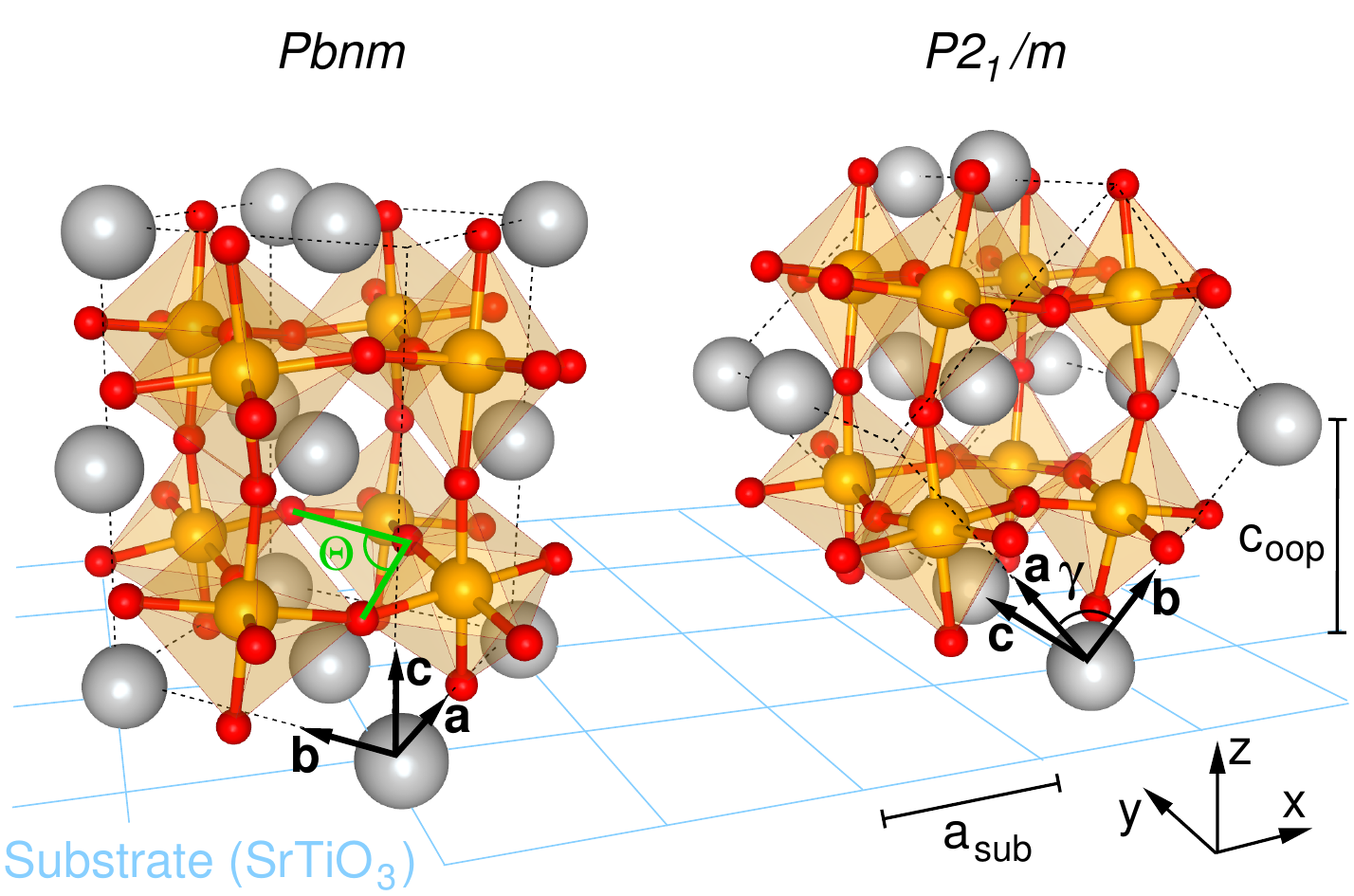}
  \caption{(Color online) Schematic depiction of two different orientations of the bulk crystal structure of \lvo\ relative to a hypothetical square lattice substrate (e.g. SrTiO$_3$), corresponding to (left) the [001] and (right) the [110] \pbnm\ growth directions.
  The respective unit-cells (dashed lines) and their basis vectors, \textbf{a}, \textbf{b}, and \textbf{c}, are indicated.
  The grey, golden, and red spheres represent La, V, and O atoms, respectively.
  }\label{fig:geom}
\end{figure}

A recent experimental study discussed the role of epitaxial strain on the electronic properties of the Mott-insulators \lto\ and \lvo\ \cite{He2012PRB}.
By growing thin films of these materials on different substrates, it was shown that a metal-insulator transition occurs in \lto\ under compressive strain, such that the whole film becomes metallic when grown on a \sto\ substrate \cite{Wong2010PRB}.
In \lvo, instead, conductivity is limited to the interface region \cite{He2012PRB,Hotta2007PRL}, suggesting that away from the interface the Mott-insulating state in \lvo\ is not affected by epitaxial strain.
Theoretical studies of strained \lto\ indeed confirm an insulator-to-metal transition
under compressive strain \cite{Dymkowski2014PRB,Ishida2008PRB}, implying that interface effects are secondary in this system.
Furthermore, recent simulations of \lto/\sto\ superlattices consistently report a metallic state in the compressively strained \lto\ layers \cite{Lechermann2013PRB}.

The appropriate technique for calculating the electronic properties of this class of correlated transition metal oxides is based on density functional theory (DFT) plus dynamical mean field theory (DMFT) \cite{Pavarini2011book,Held2007AdvP,Lechermann2006PRB}, which has been applied to study the bulk phases of both \lto\ \cite{Pavarini2004PRL,Craco2004PRB} and \lvo\ \cite{DeRaychaudhury2007PRL,Dang2014PRBr,Haule2014PRB}, as well as to address the effect of epitaxial strain in \lto\ \cite{Dymkowski2014PRB}. 
A theoretical study of epitaxially-strained \lvo\ is highly desirable in order to correctly interpret the different properties observed in thin films of \lvo\ compared to \lto\ \cite{He2012PRB}, and also to achieve a more comprehensive picture of strain effects in Mott-insulating complex oxides.

To fill this gap, we address the role of epitaxial strain on the structural and electronic properties of \lvo\ using DFT, DFT+U \cite{Cococcioni2005PRB}, and DFT+DMFT calculations. 
The effect of epitaxial strain arising from the film/substrate lattice mismatch is isolated from explicit interface effects (not addressed in this work) by modeling the thin film with a bulk-like geometry subject to properly chosen epitaxial constraints on its lattice parameters \cite{Rondinelli2011AM}.
Two different orientations of the bulk crystal structure relative to the film/substrate interface are considered here (see \pref{fig:geom}):
The first orientation corresponds to the growth along the orthorhombic [001] direction of the bulk structure.
This preserves the \pbnm\ space-group symmetry of the paramagnetic bulk phase, as well as its $a^- a^- c^+$ tilt system (in Glazer's notation \cite{Woodward1997AC}).
The second orientation corresponds to the growth along the orthorhombic [110] direction.
In this case the epitaxial constraint lowers the space group symmetry to \ptobm.
The resulting unit-cell is monoclinic with $\gamma \neq 90^\circ$ ($\gamma$ being the angle between the two short axes \bfa\ and \bfb\ in the \pbnm\ setting), and in principle allows for the more generic tilt system $a^- b^+ c^-$ \cite{Woodward1997AC}\footnote{Notice that the in-phase rotations are now along $y$ because the long axis of the parent \pbnm\ cell lies in the epitaxial $xy$ plane.
We also point out, that the \ptobm\ symmetry considered here is distinct from the low-temperature magnetically-ordered phase of \lvo, which instead corresponds to space group \ptobb\ with the monoclinic angle between the long axis and one of the short axes.}.
The [110] growth orientation with \ptobm\ symmetry and $a^- a^+ c^-$ tilt pattern was recently reported in room-temperature \mbox{x-ray} diffraction (XRD) experiments on \lvo\ thin films grown on a \sto\ substrate \citep{Rotella2012PRB,Rotella2015JPCM2}. 

Our results show no clear energetic preference toward one of the two geometries, both under compressive and tensile strain, suggesting that the actual growth orientation is mainly determined by the specific film/substrate interface and/or growth conditions.
Furthermore, the Mott-insulating phase of \lvo\ is found to be rather robust against strain and the Mott gap survives also under large compressive strains.

The paper is organized as follows.
In \pref{sec:method}, we describe the computational methods and parameters for our calculations.
In \pref{sec:geom}, the geometry and energetics of the two crystal orientations are calculated and compared as a function of epitaxial strain. We concentrate mainly on the room-temperature paramagnetic structure, but for comparison we also present some DFT+U results for a magnetically-ordered phase. 
In \pref{sec:dmft}, the electronic properties are studied as a function of strain using DFT+DMFT, based on the relaxed nonmagnetic structures of \pref{sec:geom}.
Finally, we draw our conclusions in \pref{sec:concl}.

\section{Methods and computational details}\label{sec:method}

The geometry optimizations and band structure calculations were performed within the generalized-gradient approximation (GGA) \cite{Perdew1996} of DFT using the plane waves/pseudopotential implementation from the \qe\ package \cite{QE-2009}.
The interaction between core-electrons and valence electrons is described through ultrasoft pseudopotentials \cite{Vanderbilt1990}, with the $3s$ and $3p$ ($5s$ and $5p$) semicore states of V (La) included in the valence, but without nonlocal projectors on the empty \mbox{La-$4f$} states.
The electronic wavefunctions and charge densities are expanded in a plane wave basis set with kinetic energy cutoffs of 40 Ry and 300 Ry, respectively.

\begin{table*}[t]
  \setlength{\tabcolsep}{9pt}
  \begin{tabular}{ccccccccc}
      & method&   $a$  &   $b$  &   $c$  &  \atet & \acub  &  $\theta$  &  $\phi$   \\ \hline
\multirow{2}{*}{PM} & Exp.  &  5.55  &  5.56  &  7.83  &  3.93  &  3.92  &\adeg{11.5}&\adeg{ 7.4} \\ 
                    & $U=0$ &  5.48  &  5.53  &  7.77  &  3.89  &  3.89  &\adeg{10.9}&\adeg{ 7.4} \\
  \hline
      & $U=0$ &  5.53  &  5.62  &  7.73  &  3.94  &  3.92  &\adeg{13.0}&\adeg{ 8.5} \\
G-AFM & $U=3$ &  5.57  &  5.68  &  7.83  &  3.98  &  3.96  &\adeg{15.1}&\adeg{ 9.4} \\
      & $U=5$ &  5.59  &  5.76  &  7.85  &  4.01  &  3.98  &\adeg{15.7}&\adeg{10.1} \\
  \end{tabular}
  \caption{Cell parameters (in \AA) and octahedral tilt angles for the paramagnetic (PM) and G-type antiferromagnetic (G-AFM) bulk-phase obtained within GGA+U for different $U$ (in \ev) and within plain GGA ($U=0$).
    The averaged pseudo-tetragonal and pseudo-cubic cell parameters are defined as $\atet=(a+b)/2\sqrt{2}$ and $\acub = (a+b+c/\sqrt{2})/3\sqrt{2}$. 
    The conventional rotation and tilt angles for $a^- a^- c^+$ are, respectively, $\theta = (\adeg{90}-\Theta)/2$, with $\Theta$ the O--O--O angle indicated in \pref{fig:geom}, and $\phi=(\adeg{180}-\trot{z})/2$, where $\trot{z}$ is the V--O--V bond-angle along the $z$ direction. 
    Experimental values correspond to the paramagnetic structure at 150~K, determined by XRD and neutron diffraction \cite{Bordet1993}.}
  \label{tab:cell}
\end{table*}

Bulk geometries are obtained by relaxing within \pbnm\ symmetry all atomic positions and cell parameters until forces become smaller than 2.6 meV/\AA, stress tensor components drop below 0.1 kbar, and the energy difference between successive steps stays below 0.14 meV.
In the GGA+U calculations these thresholds are taken as 10.3 meV/\AA, 0.2 kbar, and 0.68 meV, respectively, to speed-up convergence.
For the geometry optimizations under epitaxial strain, we consider the case where the substrate surface represents a square lattice with periodicity \asub, such as, e.g., the (001) surface of \sto.
As already pointed out in \pref{sec:intro}, we focus on the pure strain effect by using a bulk-like geometry, without explicitly considering a film/substrate interface.
The (hypothetical) substrate enters through the elastic boundary conditions imposed on the bulk unit cell, i.e., the in-plane lattice parameters are both set equal to the lattice constant \asub\ of the substrate, while the out-of-plane parameter, \coop, as well as all internal coordinates, are optimized for each value of \asub.

Specifically, for the [001] growth direction, the two short lattice vectors of the orthorhombic \pbnm\ unit cell both lie within the substrate plane (which is chosen to coincide with the $xy$ plane) and are constrained to be equal, with $a=b=\sqrt{2}\,\asub$.
This constraint preserves the \pbnm\ symmetry of bulk \lvo.
The long orthorhombic lattice vector is oriented perpendicular to the substrate plane and its length $c=2\coop$ is optimized by minimizing the out-of-plane component of the stress tensor.

For the [110] growth direction, the longest lattice vector of the orthorhombic \pbnm\ structure, $\mathbf{c}$, lies within the substrate plane (see \pref{fig:geom}). 
The epitaxial constraints can be formulated as $|\mathbf{c}| = |\mathbf{a}-\mathbf{b}| = 2\asub$ and $\mathbf{c} \perp (\mathbf{a}-\mathbf{b})$, i.e. the vectors $\mathbf{c}$ and $\mathbf{a}-\mathbf{b}$ form a square lattice with periodicity $2\asub$ parallel to the surface plane \cite{Eklund2009PRB}.
Due to the applied strain, the angle $\gamma$ between $\mathbf{a}$ and $\mathbf{b}$ will deviate from 90$^\circ$, leading to monoclinic symmetry with space group \ptobm\ \cite{Rotella2012PRB}. 
The three lattice vectors can then be expressed as follows:
\begin{equation*}\label{lattvecs}
  \bfa = \begin{pmatrix} -\asub + \epsilon_x \\ 0 \\ \coop \end{pmatrix},\;
  \bfb = \begin{pmatrix} \asub + \epsilon_x \\ 0 \\ \coop \end{pmatrix},\;
  \bfc = \begin{pmatrix} 0 \\ 2\asub \\ 0 \end{pmatrix},
\end{equation*}
where we have set the \bfc\ axis along the $y$ direction.
We will mostly consider the case $\epsilon_x = 0$, corresponding to $|\mathbf{a}|=|\mathbf{b}|$,  and optimize \coop\ through a series of total energy calculations, with full optimization of all internal structural degrees of freedom.
We will also briefly explore the case $\epsilon_x \neq 0$, given that $|\mathbf{a}|\neq|\mathbf{b}|$ in \lvo\ bulk \cite{Bordet1993}.

In the nonmagnetic calculations, the Brillouin zone is sampled with a $6 \times 6 \times 4$ grid of $k$-points and the electron occupations are broadened through a standard technique \cite{Methfessel1989} with a smearing parameter of 0.272 eV.
We checked that a denser $12 \times 12 \times 8$ $k$-point grid and a wave function cutoff of 60 Ry give the same bulk cell parameters within 0.05\% accuracy.
For the spin-polarized calculations within GGA+U, the smearing parameter was reduced to 0.027 eV and the $k$-point grid was correspondingly increased to $9 \times 9 \times 6$.

To obtain a low-energy \ttg-Hamiltonian for the DMFT calculations, the corresponding part of the DFT band structure is represented in a local basis set using maximally-localized Wannier functions (MLWFs) \cite{Marzari2012RMP,Mostofi2008CPC}.
The $k$-point sampling for the construction of these MLWFs is the same as in the total energy calculations, while the energy window is chosen according to the weight of the \ttg-like atomic wave functions within the DFT band structure (see \pref{fig:LVOproj} and \pref{sec:dmft}).
The resulting energy window extends from $-$1.5 eV below the Fermi level, \ef, up to about 1 eV for the unstrained structure (no entanglement with other bands) and about 1.5 eV for the $-4\%$ strain case, where the \ttg-bands are entangled with higher-lying bands around the $\Gamma$ point.
When needed, the disentanglement procedure is applied only to $k$-points within a small sphere (about 0.35~\AA$^{-1}$ of radius) around $k=(0,0,0)$, in order not to perturb the bands away from the entangled part of the band structure.

DMFT calculations are performed with help of the TRIQS package \cite{Parcollet2015CPC}, using a CT-HYB quantum impurity solver \cite{Gull2011RMP} and the Slater-Kanamori form of the interacting Hamiltonian with spin-flip and pair hopping-terms included.
We fix the Hund's exchange parameter $J=0.65\ev$ and increase the inter-orbital interaction $U'$ in order to find the critical value of $U=U'+2J$ at which the metal-to-insulator transition occurs (more details can be found in Ref.~\onlinecite{Dymkowski2014PRB}, where an equivalent setup has been used).
We used 10$^7$ Monte Carlo (MC) cycles to sample the interacting Green's function at an inverse temperature $\beta_{\rm T}=40\ev$, corresponding roughly to room temperature (200 MC steps inside a cycle, 2000 equilibration cycles).
The DMFT self-consistent cycle on the Green's function contained 30 iterations. This was sufficient to achieve good convergence of the impurity Green's function, which was expanded in a series of Legendre polynomials with 35 terms \cite{Boehnke2011PRB}.
In the \pbnm\ geometry, all four V sites are equivalent to each other by symmetry, thus only one effective impurity problem needs to be solved.
In the \ptobm\ geometry instead, two independent impurity problems have to be solved, due to the symmetry lowering that splits the four V sites into two distinct pairs of equivalent sites. 
The spectral function $A(\omega)$ is obtained by analytic continuation of the Green's function $G(\tau)$ onto the real axis using the maximum-entropy method \cite{Jarrell1996PR}.

\section{Structure and energetics}\label{sec:geom}

In this section we compare the energetics and structural evolution as function of epitaxial strain for the two possible growth directions discussed in the previous sections (see also \pref{fig:geom}).
We will hereafter refer to these two cases by their respective space-group symmetry, \pbnm\ and \ptobm.
Both symmetries will be studied within GGA in the paramagnetic phase (\pref{sec:PMstruct}) and within GGA+U in a magnetically-ordered phase (\pref{sec:MOstruct}).

\begin{figure}[tb]
  \centering
  \includegraphics[width=7.5cm]{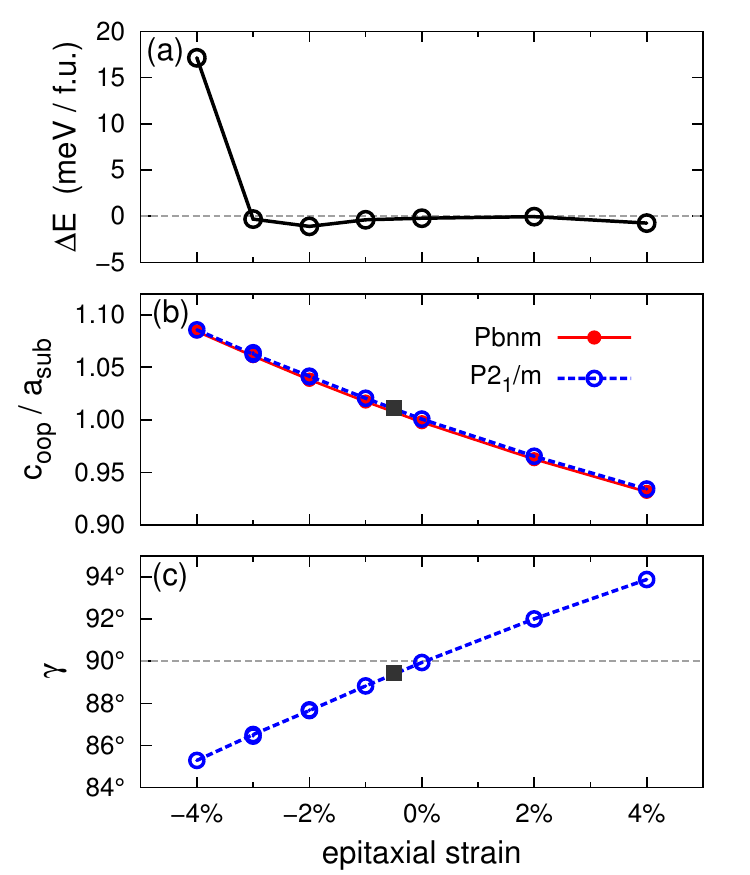}
  \caption{(Color online) (a) Total energy difference per formula unit (f.u.) between the \pbnm\ and \ptobm\ structures, $\Delta E = E_{\pbnm}-E_{\ptobm}$, (b) out-of-plane lattice parameter \coop, and (c) angle $\gamma$ between crystal axes \bfa\ and \bfb\ as functions of the imposed epitaxial strain.
  The square symbols in (b) and (c) are experimental data for epitaxial \lvo\ films on \sto\ (from Ref.~\onlinecite{Rotella2015JPCM2}).}\label{fig:struct}
\end{figure}

\begin{figure}[b]
  \centering
  \includegraphics[width=8.7cm]{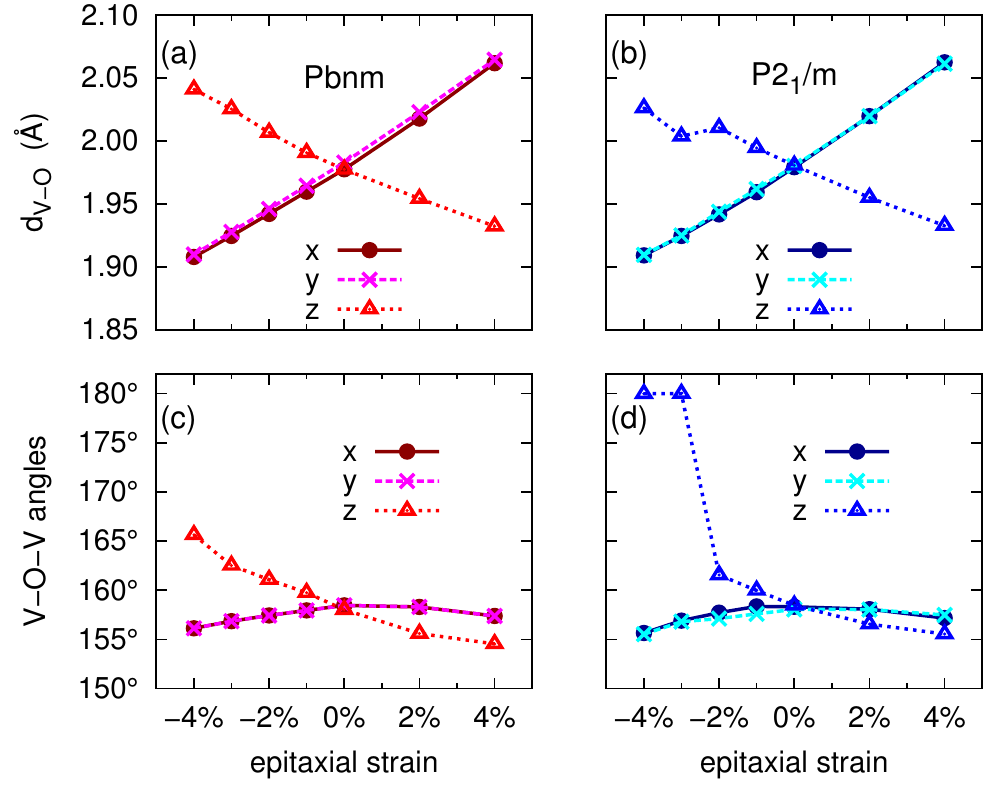}
  \caption{(Color online) (a-b) V--O bond-lengths and (c-d) V--O--V bond-angles as a function of epitaxial strain for the paramagnetic \pbnm\ and \ptobm\ geometries (on the left and on the right side, respectively).
  In the \ptobm\ case, the differences between the two nonequivalent sets of lengths and angles are small (see text), so we show only their average.}\label{fig:angles}
\end{figure}

\subsection{Paramagnetic structures}
\label{sec:PMstruct}
The relaxed \pbnm\ bulk structure obtained within spin-unpolarized GGA agrees well with the experimental structure of the paramagnetic phase at 150~K obtained from XRD \citep{Bordet1993} (see \pref{tab:cell}).
The lattice parameters $a$, $b$ and $c$ are slightly underestimated compared to the experimental data, but the deviations ($\leq 1.3\%$) are within typical limits of current DFT-based methods.
In particular, the angles $\theta$ and $\phi$, describing the strength of the octahedral tilts, are in good agreement with the experimental structure.

In the following we define the theoretical reference (i.e., zero-strain) lattice constants for epitaxially-strained \lvo\ films as $\atet=(a + b)/2\sqrt{2} = 3.891\ang$ for the \pbnm\ geometries and $\acub=(a + b + c/\sqrt{2})/3\sqrt{2} = 3.888\ang$ for the \ptobm\ case \footnote{%
For \ptobm, we use as reference the average of all three \pbnm\ bulk lattice vectors, because none of them is perpendicular to the interface in this growth geometry.
An alternative way of defining the reference lattice constant is the value of \asub\ that minimizes the total energy of the respective unit cell, corresponding to the so-called ``coherent structure'' \cite{Rondinelli2011AM}.
For both \pbnm\ and \ptobm\ geometries, we find that the optimal \asub\ differs by less than $0.01\%$ from, respectively, \atet\ and \acub\ as defined in the text.}.
We note that in the nonmagnetic DFT structure the difference between \atet\ and \acub\ is essentially negligible, as is the case in the experimental paramagnetic structure \cite{Bordet1993}. 
We then relax the out-of-plane lattice parameter \coop\ and all internal degrees of freedom for several values of strain $s$ between $-4\%$ and $4\%$, corresponding to hypothetical substrates with lattice constant $\asub= (1+s) \, \atet$ or $\asub= (1+s) \, \acub$, respectively.

The energy difference between the relaxed \pbnm\ and \ptobm\ structures as function of epitaxial strain is presented in \pref{fig:struct}a.
The energies of the two structures are the same within 1\mev/formula unit (f.u.) for all strains except for a strong compressive strain of $-4\%$, where the \ptobm\ structure is favored energetically by about 17\mev/f.u.
The out-of-plane lattice expansion (contraction) in response to the compression (extension) of the in-plane lattice constant is also virtually identical in both symmetries (see \pref{fig:struct}b).
This means that the two cells have approximately the same volume at all strains,
even if their shape differs.
Indeed, the monoclinic angle $\gamma$ of the \ptobm\ cell, shown in \pref{fig:struct}c, deviates from the \adeg{90} value of the orthorhombic \pbnm\ cell as a result of the change in the $\coop/\asub$ ratio. 
The experimental report of $\gamma\simeq\adeg{89.45}$ and $\coop/\asub \simeq 1.012$ for \lvo\ films on \sto\ \cite{Rotella2015JPCM2} closely agrees with our theoretical prediction extrapolated for the corresponding strain value (about $0.5\%$, see \pref{fig:struct}b and \pref{fig:struct}c).

The evolution of the internal structural degrees of freedom, i.e., V--O bond-lengths and V--O--V bond-angles, as function of strain is depicted in \pref{fig:angles}.
Instead of the tilt and rotation angles, $\phi$ and $\theta$, that are often used to quantify octahedral rotations in perovsites with \pbnm\ symmetry (see \pref{tab:cell} and, e.g., Refs.~\onlinecite{Dymkowski2014PRB} and \onlinecite{Rondinelli2011AM}), here we use the V--O--V bond angles, \trot{i} (where $i=x,y,z$ indicates the approximate orientation of the corresponding bonds).
This is in order to facilitate a better comparison with the \ptobm\ case, which contains three independent Glazer rotations.
We also note that within \ptobm\ symmetry there are two pairs of symmetry-equivalent octahedra per unit cell, so in principle two sets of bond-distances and four different bond-angles can be distinguished.
However, the corresponding differences are found to be minor, so we show here only the average values (distances differ by less than $0.4\%$ between the two sites; differences in \trot{y} are less than $0.5\%$, except for $s=+4\%$, where the difference is $1.3\%$).
It can be seen from \pref{fig:angles} that, similar to the cases of the total energy and \coop\ shown in \pref{fig:struct}, the strain response of the internal degrees of freedom is also very similar for the two symmetries, except for compressive strains stronger than $-2\%$.

The change in the in-plane ($xy$) V--O distances depicted in Figs.~\ref{fig:angles}a-b resembles that of the in-plane lattice parameter (as imposed by strain), while the out-of-plane ($z$) V--O bond behaves oppositely and follows the evolution of \coop\ (see \pref{fig:struct}b).
Thus, these structures do not behave like a system of rigid edge-connected octahedra and the lattice expansion/shrinkage is accommodated by changes in \emph{both} octahedral tilt-angles and V--O distances. 
Remarkably, for $s \leqslant -3\%$, the \ptobm\ structure exhibits a complete tilt-suppression along the out-of-plane direction, with a complete straightening of the V--O bonds along $z$ and the tilt pattern reducing to $a^0 a^0 c^-$.
Correspondingly, the V--O distances and the total energy difference $\Delta E$ display a sudden change, with \ptobm\ becoming more favorable for strains $s < -3\%$. 

\begin{figure}[b]
  \includegraphics[width=8.5cm]{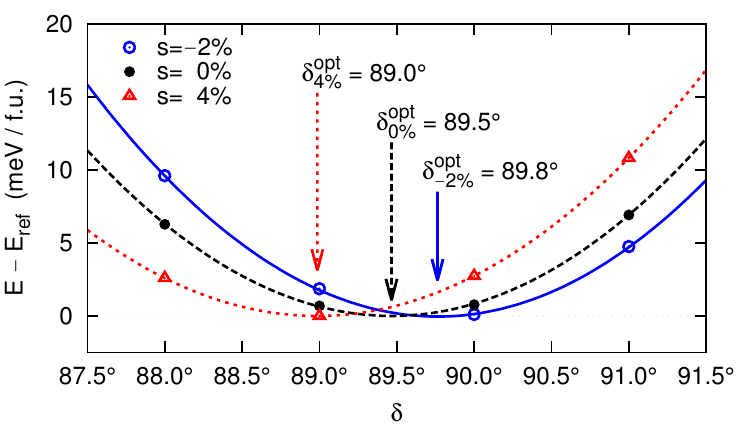}
  \caption{(Color online) Total energy $\etot(\delta)$, minimized with respect to \coop, as a function of the angle $\delta$ between the substrate and the vector $\bfa+\bfb$ in the \ptobm\ unit cell for different values of epitaxial strain $s$.
   For each $s$ value, the optimal angle $\delta^{\rm opt}_{s}$ obtained through a second order polynomial fit of $\etot(\delta)$ (lines) is shown by the arrows.
   All data and curves are shifted in energy by their respective minimum.}
  \label{fig:fitEtotaa}
\end{figure}

The changes in the tilt angles upon compressive strain from our calculations (see Figs.~\ref{fig:angles}c-d) agree with the experimental observations of \citet{Rotella2012PRB} on \lvo\ thin films grown on a \sto\ substrate, which imposes an epitaxial strain of about $-0.5\%$. 
They report enhanced octahedral rotations around the out-of-plane axis, leading to smaller in-plane V--O--V bond angles compared to the bulk, and a reduced tilting around the two in-plane directions, leading to a larger out-of-plane bond angle.
Quantitatively, the calculated departure of \trot{z} from the bulk value for strains between $0\%$ and $-1\%$ is much smaller than for the reported experimental value of $\trot{z}\simeq\adeg{171.3}$ \cite{Rotella2012PRB}, nevertheless both experiment and theory agree on the overall trends.
It should also be noted that, as pointed out in Ref.~\onlinecite{Rotella2015JPCM2}, there is a significant uncertainty in the experimental value, due to the difficulties in accurately determining oxygen positions within a thin film structure.

Finally, we also explored the possibility that the out-of-plane direction of the pseudo-cubic perovskite lattice of the \lvo\ film ($\bfa+\bfb$ in \pref{fig:geom}) might not be perpendicular to the substrate plane, resulting in $|\bfa|\neq|\bfb|$ as for the bulk \pbnm\ structure \cite{Bordet1993}.
By varying the value of $\epsilon_x$ (see \pref{sec:method}), we searched for the optimal angle $\delta$ between the $xy$ plane and $\bfa+\bfb$ for several values of epitaxial strain in the \ptobm\ symmetry.
The results, presented in \pref{fig:fitEtotaa}, show that with a perfectly matching substrate ($s=0\%$) the optimal angle deviates slightly from \adeg{90} ($\delta_{\rm opt} \simeq \adeg{89.5}$), and this deviation increases under tensile strain ($\delta_{\rm opt} \simeq \adeg{89.0}$ at $+4\%$ strain).
Under compressive strains, instead, we find that $\bfa+\bfb$ becomes almost perpendicular to the substrate plane ($\delta_{\rm opt} \simeq \adeg{89.8}$ for $s=-2\%$), supporting the experimental finding for \lvo\ thin films on \sto\ \cite{Rotella2012PRB}.

\subsection{Magnetically-ordered structures}
\label{sec:MOstruct}

Even though the bulk structure of \lvo\ obtained from our spin-unpolarized GGA calculations agrees rather well with the experimental structure obtained for the paramagnetic phase at 150\,K, it should be noted
that the corresponding calculation results in a metallic state, whereas in the experiment \lvo\ is a paramagnetic insulator.
An insulating state can be obtained within DFT+U in the presence of both magnetic order (MO) and orbital order (OO) \cite{Fang2004PRL}.
However, we point out that, in practice, neither DFT nor DFT+U can faithfully describe the paramagnetic room temperature Mott-insulating phase of \lvo.
Nevertheless, in order to cross-check the results obtained from the nonmagnetic GGA calculations, in particular the very small energy difference between the two different growth orientations (except under strong compressive strain) and the general trends of the structural parameters, we now present results of GGA+U calculations for a magnetically-ordered phase of \lvo.

\begin{figure}[t]
  \center
  \includegraphics[width=7.5cm]{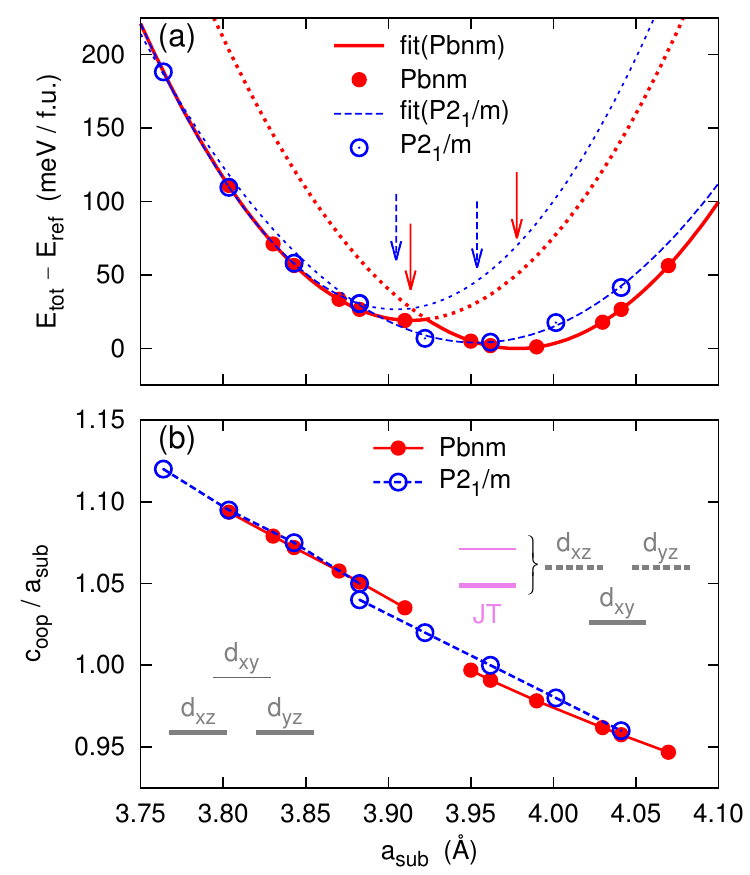}
  \caption{(Color online) (a) Total energy and (b) $\coop/\asub$ ratio as function of the substrate lattice constant \asub\ for the \pbnm\ (points) and \ptobm\ (circles) cases with G-type AFM order and $U=3\ev$.
  The lines represent polynomial fits to the data (see text) and the arrows indicate the corresponding minima.
  The reference energy $E_{\rm ref}$ is set to the lowest \pbnm\ minimum ($\asub\simeq3.98\ang$).
  Schematic representations of the strain-induced \ttg\ crystal-field splitting under compressive and tensile strain are shown as insets in (b).
  Thick (thin) solid lines represent fully-occupied (empty) levels in the presence of magnetic order, while dashed lines represent partially occupied levels that can lead to a JT distortion. } 
  \label{fig:EtotAFMG}
\end{figure}

The experimentally observed low-temperature phase of \lvo\ shows C-type anti-ferromagnetic (AFM) order, corresponding to G-type OO and a symmetry lowering to \ptobb\ space group \cite{Bordet1993,Miyasaka2003PRB,DeRaychaudhury2007PRL}.
Instead, we are here interested in the strain dependence of the room-temperature \pbnm\ phase. 
The V--O bond-length disproportionation compatible with \pbnm\ symmetry promotes C-type OO, which, according to the Anderson-Goodenough-Kanamori rules as well as experiments \cite{Miyasaka2003PRB}, in turn favors G-type AFM order.
This combination of OO/MO is observed in the low temperature \pbnm\ phase of the closely related compound YVO$_3$ \cite{Noguchi2000PRB,Fang2004PRL} and several members of the $R$VO$_3$ series, with $R$ a small rare earth cation \cite{Miyasaka2003PRB}.
Therefore, in the following we use GGA+U calculations with G-type MO in order to check the structural evolution of an insulating \pbnm\ phase of \lvo\ as function of epitaxial strain for both \pbnm- and \ptobm-type growth orientations.

\begin{figure}[b]
  \includegraphics[width=8.3cm]{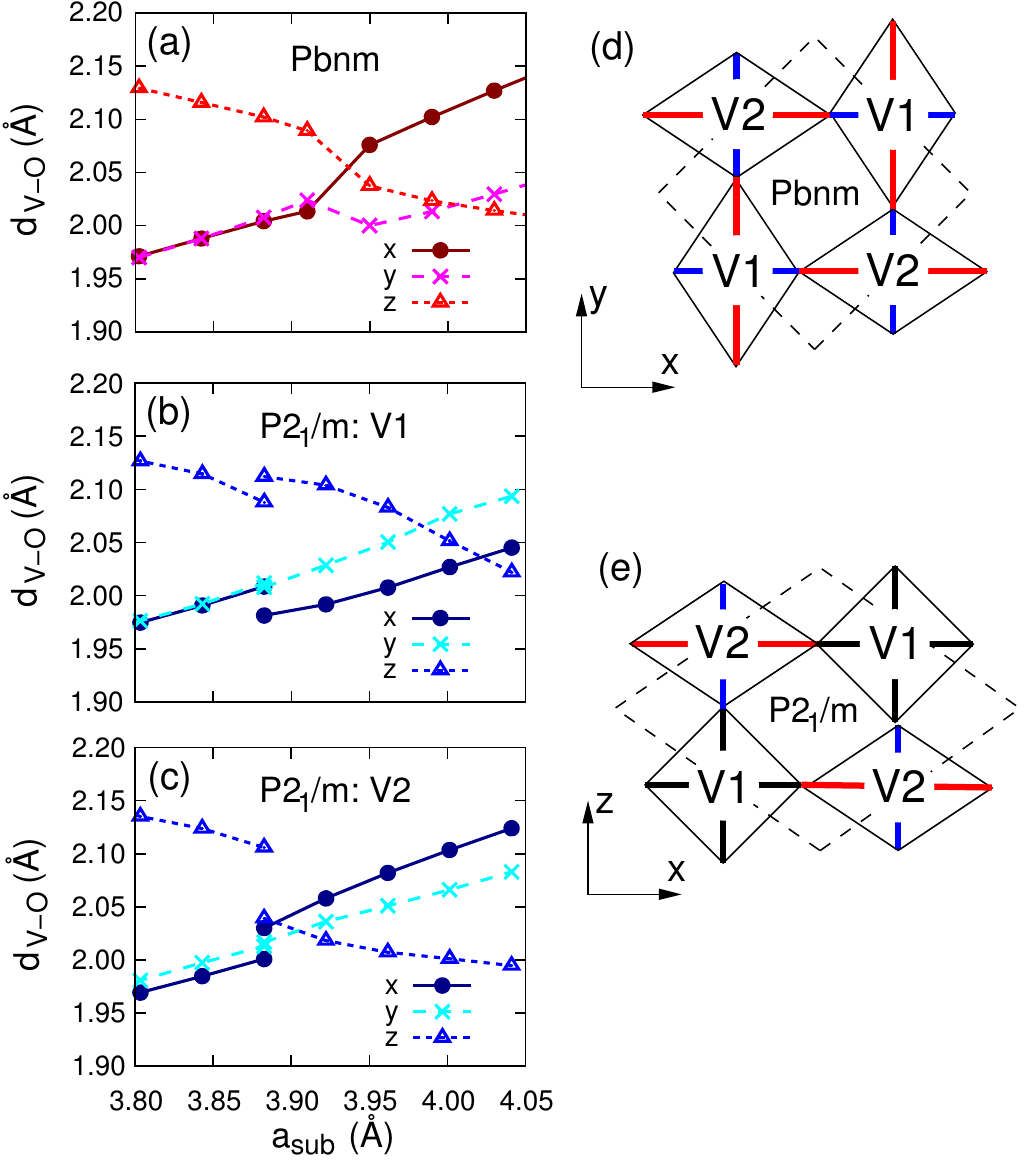}
  \caption{(Color online) V--O bond-lengths as a function of \asub\ for the \pbnm\ and \ptobm\ geometries with G-type AFM order and $U=3\ev$.
  For the \ptobm\ case, results for both inequivalent octahedra, (b) V1 and (c) V2, are shown.
  Within \pbnm\ symmetry (a), \dvo\ along $x$ and $y$ are swapped for neighboring octahedra in the $xy$ plane. 
  (d) Schematic representations of the V--O bond length pattern under tensile strain ($\asub \simeq 4.04\ang$) in the $xy$ plane of the \pbnm\ cell and (e) in the $xz$ plane of the \ptobm\ cell.
  Red (blue) segments represent long (short) V--O bonds, while the black ones have intermediate values.
  The pattern repeats equally in all planes normal to the figure because of the mirror symmetry $m$.}
  \label{fig:bondsAFMG}
\end{figure}

The optimized cell parameters and tilt/rotation angles of the unstrained bulk \pbnm\ structure with G-type AFM order for several values of $U$ are listed in \pref{tab:cell}.
Comparing first the two cases with and without MO for $U=0$, one can see that the introduction of MO leads to an overall increase in volume and a larger amplitude of the octahedral tilt distortions.
Furthermore, the orthorhombic distortion of the unit cell is more pronounced, i.e., the $b/a$ and $c/\sqrt{2}a$ ratios deviate more strongly from unity compared to the nonmagnetic case. 
Thus, while the overall volume for $U=0$ and G-AFM order agrees well with the experimental structure, the structural distortions are less well described compared to the nonmagnetic case.
A finite value of $U$ further increases the volume and the orthorhombic distortion of the unit cell, as well as the octahedral tilts, as shown in \pref{tab:cell} for $U=3\ev$ and $U=5\ev$.
In the following we will report the data for $U=3\ev$, which results in a DFT band gap of $\sim 1.3\ev$ for the bulk structure and was also used in previous DFT+U studies \cite{Fang2004PRL}.
For different strains the band gap varies by about $0.3\ev$, but all structures are insulating.

The total energy and the \coop/\asub\ ratio for the strained structures corresponding to the two different orientations are presented in \pref{fig:EtotAFMG} as a function of \asub.
We notice that the difference between the averaged lattice parameters \atet\ and \acub\ becomes non-negligible in the AFM structures, thus we compare the \pbnm\ and \ptobm\ cases as function of the substrate lattice constant, \asub, not as functon of strain $s$.
It is apparent from \pref{fig:EtotAFMG}, that the calculated energies for each orientation cannot be well described by a single low-order polynomial and that the respective \coop/\asub\ ratios exhibit a small jump at some value of \asub. This points to the presence of at least two different phases, which, as we discuss below, result from the competition between the applied strain and the Jahn-Teller (JT) distortion related to the C-type OO.

In the \pbnm\ case, the total energy (\pref{fig:EtotAFMG}a) can be well fitted by two parabolas intersecting around $\asub=3.92\ang$, where the \coop/\asub\ ratio displays a small jump (\pref{fig:EtotAFMG}b).
For $\asub<3.92\ang$, the strain-induced crystal-field splitting (see inset in \pref{fig:EtotAFMG}b) together with the local spin-splitting on the V sites is sufficient to create an insulating state by occupying the two lowest spin-(local)-majority \ttg-levels (\dxz\ and \dyz), without requiring a further structural distortion.
In contrast, the crystal-field splitting under tensile strain requires an additional JT distortion to lift the remaining orbital degeneracy and open up a gap between completely filled and completely empty states (see inset in \pref{fig:EtotAFMG}b).
Therefore, above a certain critical value of \asub\ the system exhibits a staggered JT distortion of the V--O bond lengths (see \pref{fig:bondsAFMG}d) corresponding to the C-type OO imposed by the \pbnm\ symmetry in combination with the G-type MO.
Here, the ``staggering'' related to the C-type OO occurs within the $xy$ plane, i.e., parallel to the hypothetical substrate surface, see also \pref{fig:bondsAFMG}d, so that both directions within this plane experience the same epitaxial constraint.
This picture is confirmed by the evolution of the V--O bond distances with strain, shown in \pref{fig:bondsAFMG}a, where a bond disproportionation in the $xy$ plane is clearly visible for $\asub>3.92\ang$.
Notice, however, that such large bond disproportionation is not present in the paramagnetic bulk phase \cite{Bordet1993}.

The \ptobm\ case is slightly more complex, but can be interpreted along the same lines.
For strong enough compressive strain ($\asub<3.88\ang$), the strain-induced crystal-field splitting suffices to open a gap and an additional JT distortion is absent (see Figs.~\ref{fig:bondsAFMG}b-c), analogously to the \pbnm\ case.
For $\asub \ge 3.88\ang$ a transition to a more complex JT-distorted phase can be observed (both phases can be stabilized for $\asub\simeq3.88\ang$).
In contrast to the \pbnm\ case, here the staggering plane ($xz$) is perpendicular to the substrate plane ($xy$).
Therefore, the short-bond/long-bond alternation within the $xz$ plane is superimposed to opposing trends governed by the epitaxial constraint (along $x$) and the elastic response (along $z$). 
As a result, the V--O bond distances within the $xz$ plane become nearly equal for the V1-site under tensile strain and for the V2-site under compressive strain (the first case is depicted in \pref{fig:bondsAFMG}e), whereas the corresponding bond lengths for the respectively other V site are very asymmetric.

The fits to the total energy for the \pbnm\ and \ptobm\ geometries (see \pref{fig:EtotAFMG}a) show that the corresponding energy difference is very small for any $\asub<3.90\ang$, while the \ptobm\ orientation is slightly favored over \pbnm\ within a tiny range around $\asub\sim 3.92\ang$, but then becomes unfavorable for $\asub>3.96\ang$.
Overall, the energy difference between \pbnm\ and \ptobm\ always stays below 10 meV/f.u. Furthermore, the small energy differences visible in \pref{fig:EtotAFMG}a seem to be closely related to the JT distortion, which is absent in the room temperature structure of \lvo.
Therefore, no clear preference toward one of the two geometries can be concluded based on the calculated bulk-like elastic response.
This confirms the corresponding result obtained for the case without MO (see the preceeding section), and suggests that the specific orientation of the \lvo\ unit cell on the subtrate is determined mostly by other factors, such as effects related to the specific substrate/film interface or the details of the growth process.

\begin{figure}[tb]
  \includegraphics[width=8.3cm]{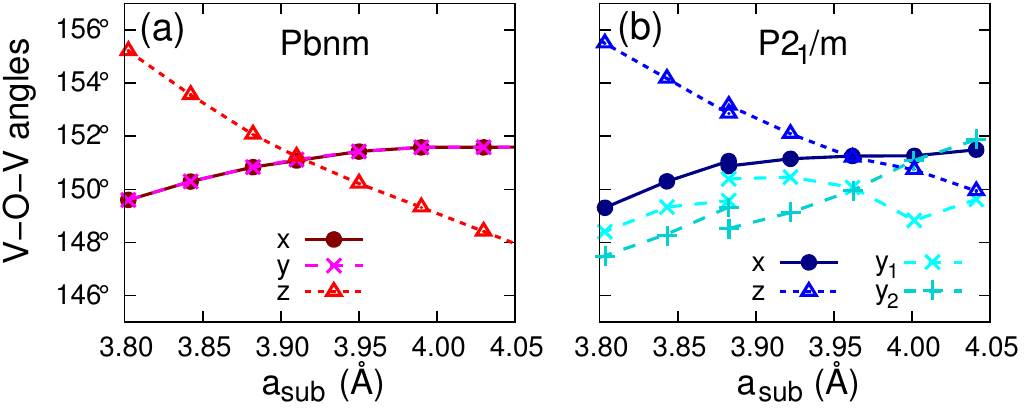}
  \caption{(Color online) V--O--V bond-angles as a function of \asub\ for (a) the \pbnm\ and (b) the \ptobm\ geometries with G-type AFM order and $U=3\ev$.
  For the \ptobm\ case, both \trot{y_1} (between V1-type octahedra stacked along $y$, see \pref{fig:bondsAFMG}e) and \trot{y_2} (between V2-type octahedra) are shown.}
  \label{fig:anglesAFMG}
\end{figure}

Finally, the V--O--V bond-angles, shown in \pref{fig:anglesAFMG}, also exhibit similar trends as in the nonmagnetic DFT calculations (\pref{fig:angles}). 
However, due to the overestimation of the octahedral tilt distortions (see above), the bond angles deviate more strongly from the ideal \adeg{180} compared to the case without MO. 
Consequently, the \ptobm\ geometry does not show the full suppression of tilts around the in-plane directions found in the nonmagnetic case under strong compressive strain.
Nevertheless, the increasing value of \trot{z} with compressive strain shows the same tendency for straightening the out-of-plane bonds as observed in the thin film experiments \cite{Rotella2012PRB}.

\section{Electronic structure within DFT+DMFT}\label{sec:dmft}

\begin{figure}[t]
  \includegraphics[width=8.3cm]{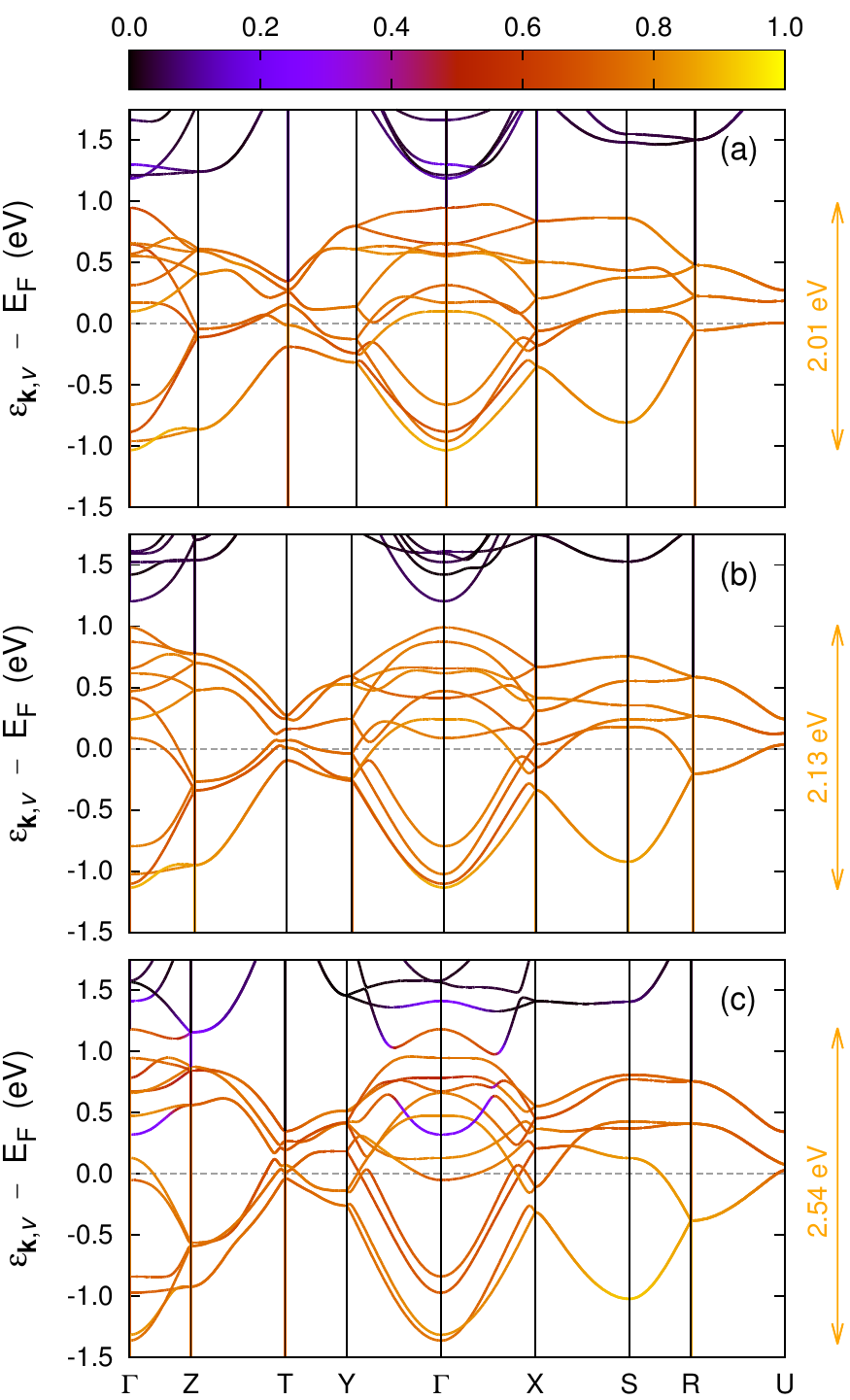}
  \caption{(Color online) Band structure close to the Fermi level for epitaxially-strained \lvo\ under (a) tensile strain ($s=4\%$) (b) $0\%$ strain, and (c) compressive strain ($s=-4\%$) in the \pbnm\ growth geometry.
  For each Kohn-Sham eigenvalue $\varepsilon_{\bfk,v}$, the weight of the V-\ttg\ orbital character in the corresponding wave function is represented on the color scale shown above (a).
  The width of the \ttg\ group of bands is also shown for each strain value.}
  \label{fig:LVOproj}
\end{figure}

Next we address the electronic properties of epitaxially strained \lvo, in particular the effect of strain on the insulating character of the paramagnetic phase observed in the bulk system above 140 K \cite{Miyasaka2000PRL}.
As already pointed out at the beginning of \pref{sec:MOstruct}, none of the available DFT approximations can correctly describe this Mott-insulating phase, therefore we will work in the more appropriate theoretical framework of DFT+DMFT.
In the absence of magnetic order, the DFT-GGA band structure of \lvo\ (shown in \pref{fig:LVOproj}) presents a group of bands around the Fermi energy and no energy gap.
This group of bands is rather well-isolated from other bands at lower and higher energies. By projecting the Kohn-Sham states onto atomic-like V-$3d$ wave functions with \ttg\ symmetry (i.e., \dxz, \dyz, and \dxy), one can verify that this group of bands consists predominantly of V-\ttg\ orbital states (see \pref{fig:LVOproj}), hybridizing with the $2p$ states of the neighboring oxygen atoms.
The V-\ttg\ weight elsewhere in the band structure is always below $20\%$ and concentrated mostly on the O-$2p$-related bands located below \ef\ (between about $-4\ev$ and $-8\ev$, not shown).

Most of the low energy physics of \lvo\ is determined by these partially filled bands with dominant V-\ttg\ character present around the Fermi level \cite{DeRaychaudhury2007PRL,Dang2014PRB}.
We therefore construct a basis of MLWFs for these ``V-\ttg\ bands'' (details in \pref{sec:method}), starting from initial projections of the Bloch functions in the corresponding energy window on atomic V-\ttg\ orbitals.
This basis is then used to represent the low-energy non-interacting Hamiltonian in our DMFT calculations.
An explicit inclusion of the O-$p$ bands, which are separated by a gap of about $2.5\ev$ from the V-\ttg\ bands, would not have a significant effect on the obtained results \cite{Dang2014PRB,Haule2014PRB}.
All DMFT calculations are performed for the epitaxially-strained structures obtained using spin-unpolarized GGA, already discussed in \pref{sec:PMstruct}.

\begin{figure}[tb]
  \centering
  \includegraphics[width=6.9cm]{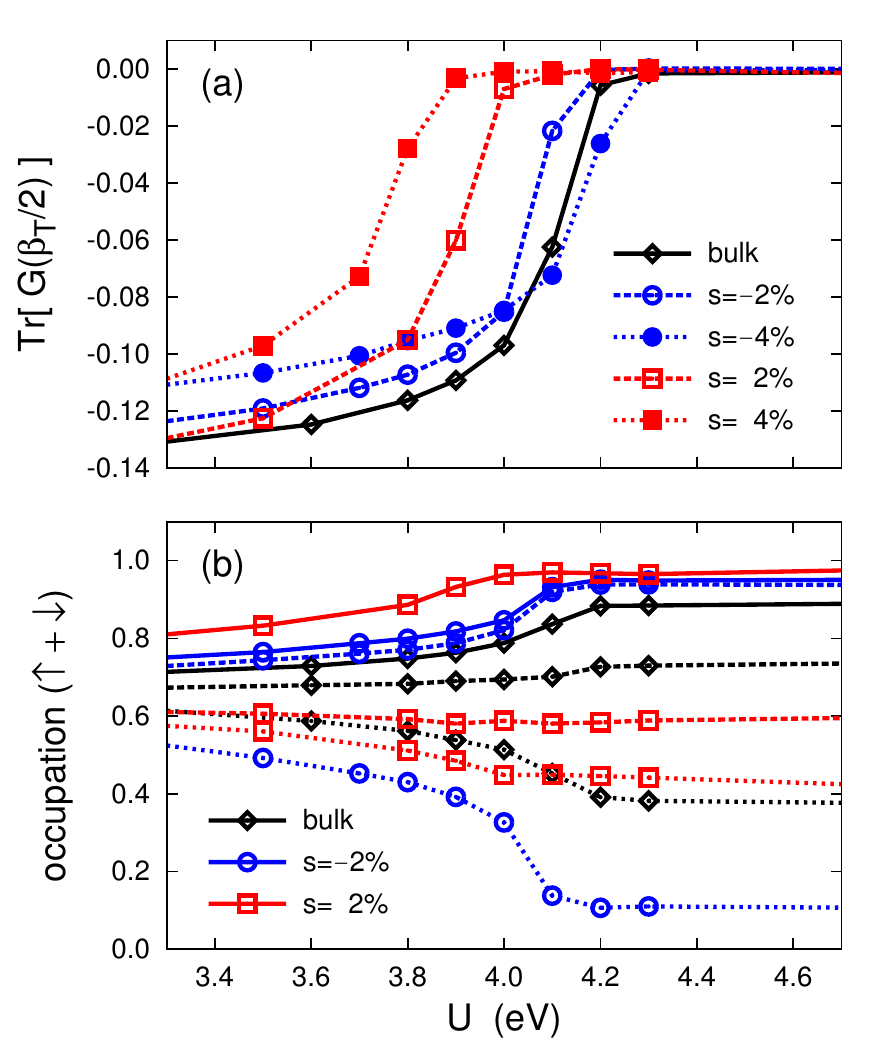}
  \caption{(Color online) DMFT results for bulk \lvo\ (black) and for \lvo\ under tensile (red) or compressive (blue) epitaxial strain within the \pbnm\ geometry.
  (a) Trace of the Green's function $G(\tau)$ at $\beta_{\rm T}/2$ and (b) orbital-resolved occupations, summed over both spin components, obtained from the eigenvalues of $-G(\beta_{\rm T})$ as a function of the interaction parameter $U$.
  Different line-styles in (b) (solid, dashed, dotted) indicate different orbitals.}\label{fig:GhB1}
\end{figure}

\begin{figure}[b]
  \centering
  \includegraphics[width=8.1cm]{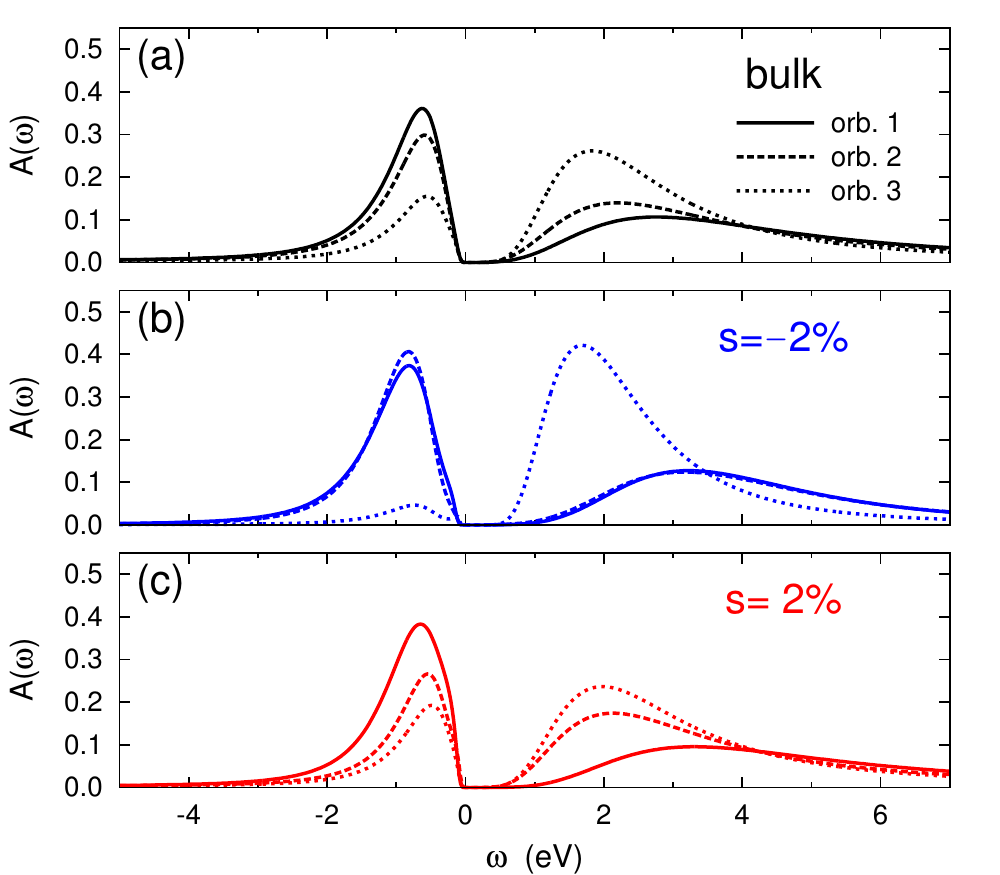}
  \caption{(Color online) Orbital components of the spectral function $A(\omega)$ obtained from DMFT for \pbnm-\lvo\ (a) in the bulk phase, (b) at $2\%$ compressive strain, and (c) at $2\%$ tensile strain computed with $U=4.8\ev$.
  The three components represent the orbital contributions within a basis set that diagonalizes the local Green's function matrix at $\tau=\beta_T$.}\label{fig:Aw}
\end{figure}

We first focus on the results obtained for the \pbnm-type growth orientation.
In \pref{fig:GhB1}a, we present the evolution of the \lvo\ spectral density at the Fermi level, $A(\omega = 0)$, as function of the interaction parameter $U$ for different strain values.
We make use of the approximate relation $\mathrm{Tr}[A_{ij}(\omega=0)] \simeq -\frac{\beta_T}{\pi}\mathrm{Tr}[G_{ij}(\beta_{\rm T}/2)]$, where $G_{ij}(\tau)$ is the impurity Green's function matrix and $i,j$ index the three \ttg-orbitals.
Above a critical value of $U$ (which we call here \umit), the spectral weight $\mathrm{Tr}[A_{ij}(\omega=0)]$ drops to zero and the system is insulating for any $U>\umit$.
In the bulk \pbnm\ system, this metal-insulator transition takes place at $\umit\simeq4.2\ev$.
The unstrained epitaxial geometry does not differ substantially from the bulk structure, and indeed we do not find any noticeable difference in $\mathrm{Tr}[G_{ij}(\beta_{\rm T}/2)]$ as function of $U$ between these two cases ($0\%$ strain case is not shown here).

When tensile strain is applied, the critical value of $U$ at which the metal-insulator transition takes place is shifted to lower values.
The system would thus reinforce its insulating behaviour when grown on a hypothetical substrate imposing tensile strain, as also found previously for \lto\ \cite{Dymkowski2014PRB}.
In the case of compressive strain, \lvo\ shows first a slight lowering of \umit\ (see \pref{fig:GhB1}a for $s=-2\%$) and then a small increase at larger strains, but the strain-induced changes are rather small.
This is at variance with \lto, where compressive strain shifts \umit\ to significantly larger values and leads to an insulator-to-metal transition already at moderate strain values of $s \sim -2\%$ \cite{Dymkowski2014PRB}.

From \pref{fig:LVOproj} it can be seen that the Kohn-Sham bandwidth of the V-\ttg\ bands increases significantly with compressive strain.
In the MLWF representation, which is relevant for the DMFT calculations, we find that the \ttg-bandwidth increases from $\sim\,$2.13 eV at $0\%$ strain to $\sim\,$2.37 eV at $-4\%$ strain (the small discrepancy of the latter with the Kohn-Sham value is related to the entanglement at the $\Gamma$-point, see \pref{fig:LVOproj}c).
It might therefore appear surprising that \umit\ is nearly unaffected by this bandwidth increase.

Further insight can be obtained by examining how the V-\ttg\ orbital occupancies change under epitaxial strain.
In \pref{fig:GhB1}b, we present the orbital occupancies $n_i\; (i=1,2,3)$, obtained from the eigenvalues of $-G_{ij}(\beta_{\rm T})$ as a function of the interaction $U$ for different values of strain.
In the bulk case (black lines), orbital polarization is very weak in the non-interacting DFT limit ($U \to 0$), and it is only slightly enhanced in the insulating phase.
For $U$ values above \umit, the three orbital occupancies (summed over both spin directions) are $n_1=0.89$, $n_2=0.75$, and $n_3=0.36$, showing that orbital fluctuations are rather strong in \lvo\ down to room temperature, consistent with what has been found in Ref.~\onlinecite{DeRaychaudhury2007PRL}.
Compressive strain suppresses these fluctuations and favors half-filling (i.e., $n_i = 1$) of two orbitals, leaving the third one essentially empty, as can be seen by the changes in the \ttg\ occupancies: $n_1 \simeq n_2 = 0.95$, $n_3=0.10$ for $s=-2\%$ (see \pref{fig:GhB1}b), and $n_1 \simeq n_2 = 0.99$, $n_3=0.02$ for $s=-4\%$.
These occupation numbers are in line with the expected tetragonal component of the crystal-field splitting under compressive strain, which lowers the energy of the $d_{xz}$ and $d_{yz}$ orbitals and raises the energy of the $d_{xy}$ orbital (see, e.g., Ref.~\onlinecite{Dymkowski2014PRB} and inset in \pref{fig:EtotAFMG}b).
The suppression of orbital fluctuations, as well as the resulting half-filling of orbitals 1 and 2, are expected to favor the insulating state and seem to counteract the bandwidth increase under compressive strain.

In contrast, tensile strain does not lead to a similar suppression of orbital fluctuations, since two orbitals are still away from half-filling ($n_2=0.60$ and $n_3=0.42$ for $s=2\%$), even if the occupation of the remaining \ttg\ orbital is now very close to one ($n_1=0.98$, see \pref{fig:GhB1}b). 
Again, these occupations reflect the expected crystal-field splitting under tensile strain (see inset in \pref{fig:EtotAFMG}b), with the in-plane (\dxy-like) orbital at lower energy than the two almost degenerate out-of-plane (\dxz- and \dyz-like) orbitals.

The full spectral functions $A(\omega)$ of \lvo\ in the presence of epitaxial strain, shown in \pref{fig:Aw}, confirm the robustness of the Mott-insulating state, as well as the suppression of orbital fluctuations under compressive strain.
The orbitally-resolved $A(\omega)$ calculated for the bulk structure using $U=4.8\ev$ leads to a band gap of $\sim\,$1\:eV (see \pref{fig:Aw}a), in good agreement with the experimentally determined value \cite{Arima1993PRB}.
Similar $U$ values were used in previous studies of \lvo\ \cite{DeRaychaudhury2007PRL,Dang2014PRB}.
Under compressive (\pref{fig:Aw}b) and tensile (\pref{fig:Aw}c) strains of $\pm2\%$ the band gap does not change noticeably and the system is still clearly insulating.
In particular, the band gap persists even for a larger compressive strain of $s=-4\%$ (not shown), consistently with the behavior of $\mathrm{Tr}[G_{ij}(\beta_T/2)]$ in \pref{fig:GhB1}a.
Under compressive strain one can notice that already for $s=-2\%$ the orbital polarization is much larger than in the bulk case, since one orbital is almost empty (short-dashed lines), while the other two are nearly half-filled (\pref{fig:Aw}b).

\begin{figure}[t]
  \centering
  \includegraphics[width=6.6cm]{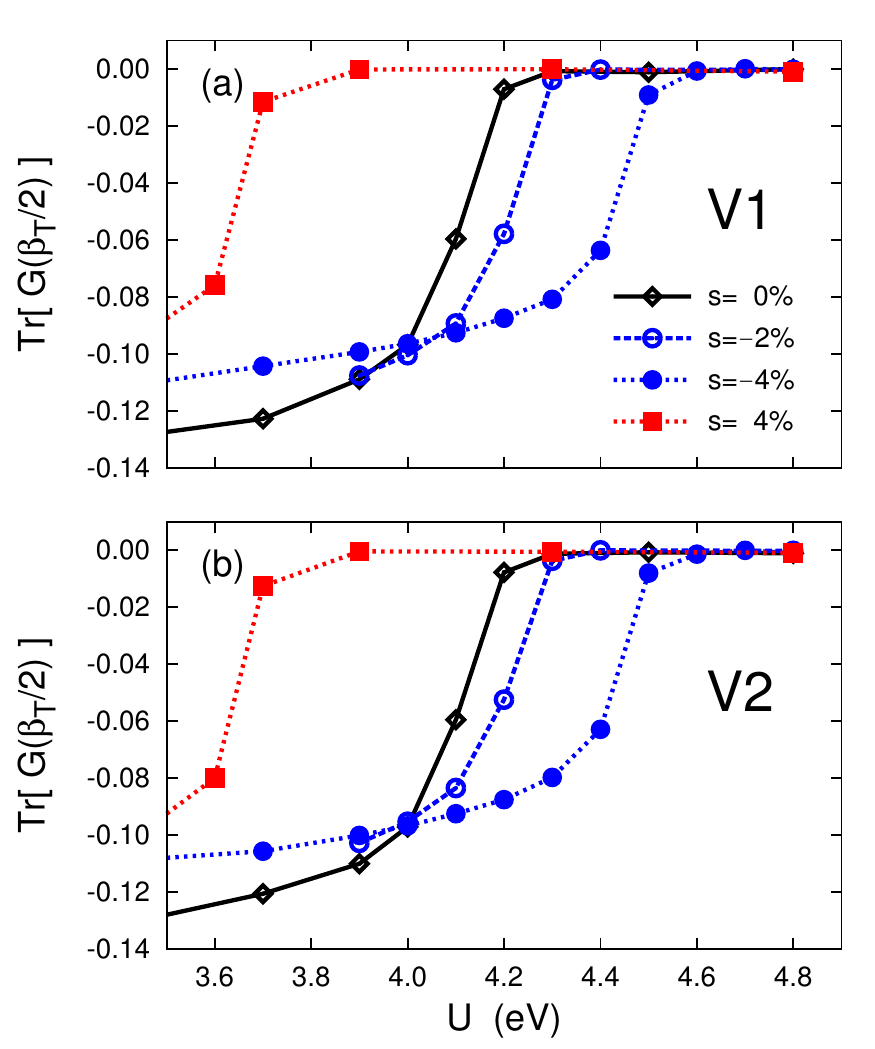}
  \caption{(Color online) Trace of the Green's function at $\beta_{\rm T}/2$ for epitaxially-strained \lvo\ in the \ptobm\ geometry for several values of the epitaxial strain.
  The two panels refer to the two inequivalent V sites, namely (a) V1 and (b) V2.}\label{fig:GhB2}
\end{figure}

In order to assess whether a different growth geometry can result into a different strain dependence in the electronic properties of \lvo, we now present results of DMFT calculations obtained for the epitaxially-strained structures with \ptobm-type growth orientation.
In \pref{fig:GhB2} we report $\mathrm{Tr}[G_{ij}(\beta_{\rm T}/2)]$ as a function of $U$ for the two inequivalent V-sites and different values of epitaxial strain for the \ptobm\ symmetry.
We note that for all $U$ values larger than 3.5~eV, the two sites show very similar behaviour and remain nearly equivalent, while for smaller values of $U$ we find a tendency for charge disproportionation between the two types of V-sites.
The disproportionation occurs for values of the interaction $U<3.8\ev$ that are well below the most reasonable range of $U=4.5\ev - 5\ev$ and indeed was never reported in experiments, thus we will not further address this issue here. 

For the \ptobm\ geometry, we find that \umit\ increases monotonously under compressive strain, reaching $U \sim 4.5\ev$ for $s=-4\%$, where a complete suppression of the octahedral tilts around the in-plane axes takes place (see \pref{sec:PMstruct}).
Because of this tilt suppression, the effect of strain on the \ttg\ crystal-field splitting is considerably reduced with respect to the \pbnm\ case.
For instance, at $s=-4\%$ the splitting between the two lower \ttg\ levels, \dxz\ and \dyz (now exactly degenerate due to the higher symmetry), and the higher-lying \dxy\ level is only 0.12 eV in \ptobm, against 0.20 eV in \pbnm.
The smaller crystal-field splitting results into a less marked orbital polarization, as can be seen from the \ttg\ occupancies calculated within DMFT: $n_1 \simeq n_2 = 0.99$, and $n_3=0.02$ for \pbnm\ (see also above), against $n_1 = 0.91$, $n_2 = 0.90$, and $n_3=0.19$ for \ptobm\ (here, we average over the two inequivalent V sites, V1 and V2, since the corresponding occupations differ by less than 0.02).
Thus, the \ttg-bandwidth increase produced by compressively straining the \ptobm\ geometry (from 2.12 eV at $s=0\%$, to 2.32 eV at $s=-4\%$ in the MLWF representation) is more effective in driving the system towards a metallic state compared to the \pbnm\ case, which shows a much stronger suppression of the orbital fluctuations.
However, the realistic value of $U\sim 4.8\ev$ is still larger than \umit\ for the strongest compressive strain considered here, $s=-4\%$, which in turn is much larger than the strain experienced by \lvo\ thin films coherently grown on \sto\ ($s \simeq -0.5\%$).
Therefore, we do not expect that the moderate strain of the \sto\ substrate could induce a metallic state in the \lvo\ film, in particular since a complete tilt suppression is not observed in this case (see also Ref.~\onlinecite{Rotella2012PRB}).
Nonetheless, noticeable changes in the spectral features, signaling the vicinity of a metallic state, could be observable for \lvo\ thin films grown on substrates that impose a larger compressive strain, such as e.g. LaAlO$_3$ (lattice mismatch of $3.3\%$).

\section{Conclusions}\label{sec:concl}

In this work, we have studied the structural response of \lvo\ under epitaxial strain, using both DFT (for the nonmagnetic/paramagnetic case) and DFT+U (for magnetically-ordered structures).
Furthermore, we have studied the corresponding evolution of the paramagnetic Mott-insulating phase using DFT+DMFT, in particular the vicinity of the material to a metal-insulator transition.
Two different symmetries, corresponding to different growth orientations of the thin film have been considered.
First, growth along the [001] direction of the bulk \pbnm\ structure, conserving the \pbnm\ symmetry, and second, growth along the [110] direction, resulting in \ptobm\ symmetry, as recently reported in thin film experiments \cite{Rotella2012PRB,Rotella2015JPCM2}.

Both for the nonmagnetic and for the magnetically-ordered structures, the two symmetries do not exhibit significant energy differences, except under strong compressive strains ($s<-3\%$) in the nonmagnetic structure.
In that case, the \ptobm\ orientation is favored by $\sim 20$ meV/f.u., as consequence of a complete straightening of the out-of-plane V--O--V bonds.
If the system is treated using DFT+U and imposing G-type AFM order to obtain a gap in the electronic band structure, the system still shows a strong tendency to straighten the out-of-plane V--O--V bonds under compressive strain, but no complete suppression of octahedral tilts occurs.
This difference can be attributed to an overestimation of octahedral tilts in the unstrained structure within G-type AFM and DFT+U.
Overall, the absolute energy difference between the \pbnm\ and \ptobm\ geometries with magnetic order stays below 10 meV/f.u.\ in a wide range of substrate lattice spacings.
Therefore, no clear preference for one or the other growth orientation is obtained from the bulk-like elastic response under epitaxial strain.
Our calculations thus suggest that the experimentally-observed occurrence of the \ptobm\ orientation in epitaxial \lvo\ films grown on \sto\ substrates \cite{Rotella2012PRB} is driven mostly by the specific substrate/film interface configuration. 

Our DFT+DMFT calculations for the strained structures show that the insulating character of \lvo\ is rather robust against epitaxial strain, and that the system is not expected to undergo any strain-driven metal-insulator transition, except maybe for very strong compressive strain ($s<-4\%$) in the \ptobm\ configuration.
This is consistent with recent experimental reports, which find that conductivity is restricted to the interface region in \lvo\ films grown on \sto\ \cite{He2012PRB}.
However, even a moderate compressive strain of $-2\%$ has the effect of suppressing orbital fluctuations (see occupations from diagonalization of DMFT density matrix), while in the Mott-insulating bulk phase of \lvo\ such fluctuations are sizable even at room temperature, in agreement with \citet{DeRaychaudhury2007PRL}.
This effect can be viewed as counteracting the increase of the \ttg\ bandwidth under compressive strain, thus impeding the formation of a metallic state.

Our study is relevant to the ongoing investigations of metal-insulator transitions in thin films of correlated transition metal-oxides, in particular for the understanding of emerging properties in complex oxide heterostructures. 
Since the origin of the metallic state in \lvo/\sto\ heterostructures is still debated \cite{Hotta2007PRL,He2012PRB,Rotella2015JPCM1}, our results provide a useful reference to further theoretical investigations on the role of other mechanism beyond strain (such as, e.g., electronic reconstructions, formation of defects, quantum confinement) that might occur at the interfaces between different oxides. 

\begin{acknowledgments}
The authors are grateful to Krzysztof Dymkowski for supplying the Wannier90/TRIQS interface.
This work was funded by ETH Zurich and the Swiss National Science Foundation through NCCR-MARVEL.
All calculations have been performed on the PASC cluster ``M\"onch'', hosted by the Swiss National Supercomputing Centre.
\end{acknowledgments}

%

\end{document}